# When web apps heal themselves: a MAPE-K based approach to fault tolerance and adaptive recovery

**Sales G. Aribe Jr., Rov Japheth G. Oracion**
Information Technology Department, Bukidnon State University, Malaybalay City, Philippines

**ABSTRACT**

Ensuring the reliability and resilience of modern web applications remains a critical challenge due to increasing system complexity and dynamic runtime environments. This study proposes a modular self-healing framework based on the monitor–analyze–plan–execute over a shared knowledge base (MAPE-K) model, integrated with an AutoFix-inspired mechanism for adaptive fault recovery. Using a design and development research (DDR) approach, the system was implemented and evaluated through controlled fault injection experiments across twenty runtime failure scenarios, including service crashes, memory leaks, and database disconnections. Experimental results demonstrate that the proposed framework achieved a mean fault detection F1-score of 90.7% and a recovery success rate of 93.2%. The AutoFix module reduced the average time-to-recovery (TTR) by 56.2%, achieving an average recovery time of 3.92 seconds. System throughput was maintained between 88% and 95% during fault conditions, with only a 3.1% increase in response time. Additionally, iterative feedback mechanisms improved recovery efficiency by 18.6% over multiple cycles. These findings indicate that the proposed framework provides a practical and extensible approach to enhancing fault tolerance in web applications through feedback-driven adaptation. While the current implementation relies on predefined recovery strategies, the integration of learning-oriented feedback establishes a foundation for future development of more autonomous self-healing systems.

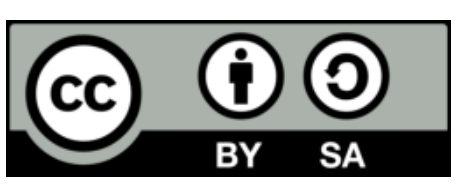

***Corresponding Author:***

Sales G. Aribe Jr.
Information Technology Department, Bukidnon State University
Malaybalay City, Philippines
Email: sg.aribe@buksu.edu.ph

## 1. INTRODUCTION

The increasing complexity of modern web applications and containerized service environments presents significant challenges in ensuring system availability, fault tolerance, and operational resilience. As deployment environments become more dynamic and user expectations for continuous service rise, traditional reactive or manual fault management approaches often prove inadequate. In response, self-healing system, a software capable of autonomously detecting, diagnosing, and repairing faults, has emerged as a promising paradigm to reduce human intervention, limit downtime, and sustain service availability [1].

Self-healing systems originate from the broader vision of autonomic computing, where systems manage their behavior through continuous feedback loops. In web application contexts, this translates into mechanisms capable of restarting failed services, rolling back to stable versions, or dynamically reconfiguring system components when faults are encountered [2], [3]. Modern orchestration tools like

Kubernetes already include basic self-healing features, such as health checks and container restarts, yet their scope remains infrastructure-bound and dependent on predefined thresholds [4], [5].

Emerging technologies across various domains demonstrate how adaptive, context-aware systems can drive digital transformation and operational efficiency. Local and international initiatives, from agricultural decision support [6] and power advisory tools [7] to digital archiving [8] and cloud-based learning platforms [9], have shown that systems capable of automated response, intelligent tuning, and self-management significantly enhance performance, reliability, and user experience [10]. Collectively, these efforts highlight that adaptive behavior and autonomous fault handling are essential features of modern software design, reinforcing the importance of building resilient and self-managing system architectures.

More advanced methods have been studied in research. For example, AutoFix is a system designed for automatically repairing programs and demonstrates how runtime faults can be fixed without involving a developer [11]. It inspects the intended behavior of a program and generates repair patches that can fix it accordingly [12].

However, existing research primarily focuses on infrastructure-level fault recovery, leaving a gap in developing self-healing capabilities that operate at the application logic level, where runtime faults such as code errors, logic failures, and service dependencies often occur. Most existing solutions rely on static rule-based fault management, which limits adaptability in unpredictable environments [13]. Prior works seldom incorporate runtime learning and feedback that could continuously enhance fault diagnosis and recovery performance [14], [15]. Related work such as SHoWA and eCrash have investigated statistical fault modelsh and service monitoring can enable autonomous recovery [16]. On the other hand, cloud-based tools such as Dynatrace and AWS CloudWatch do provide anomaly detection capabilities, but often require manual interventions, or are too dependent on static rules [17]. These lack of these features causes them to fail to instantly handle complex or unknown breakdowns. To address these gaps, a modular self-healing framework is presented in this study. It also introduces AutoFix-like intelligence with real-time monitoring, adaptive anomaly detection, and corrective actions directly in the app. The framework integrates fault localization, using mature rule-based and machine-learnt remediation techniques, and provides a closed feedback loop that learns from previous invocations and adjusts future recovery to reduce downtime.

The proposed system extends existing approaches by incorporating feedback-driven refinement within the monitor analyze plan execute–knowledge (MAPE-K) loop, enabling incremental improvement of recovery decisions over time. The core element of this system is its AutoFix-enhanced correction engine which generates corrective actions through the assessment of system state deviations from baseline conditions. The system design contains four main modules which perform behavior profiling, fault diagnosis, patch application and learning from feedback.

The system evaluation process included multiple controlled fault injection and recovery tests across different scenarios. The system's performance was evaluated through three key metrics which included anomaly detection accuracy and fault recovery time and system throughput. The system achieves two benefits through its findings by decreasing recovery delays and developing improved fault response methods which enhance operational resilience and fault tolerance. This study aims to achieve the following objectives:

- Determine fault detection accuracy in terms of precision, recall, and F1-score across various runtime failure scenarios.
- Compare automated and manual recovery performance in terms of time-to-recovery (TTR), speed improvement, and success rate using the AutoFix module.
- Evaluate the effect of the system on user experience and throughput during active fault injection under concurrent user conditions.
- Determine the effectiveness of feedback-enhanced adaptation through iterative learning within the MAPE-K loop, emphasizing on refining recovery decisions and long-term system improvement.

This study's stated objectives are to provide a realistic and adaptive approach for developing resilient web applications. Such applications would require little human intervention and would continue to function even in the face of unexpected failures.

This study contributes a modular and extensible framework for improving fault tolerance and recovery efficiency in modern web applications. It demonstrates how self-healing web frameworks, when combined with runtime feedback, behavioral learning, and AutoFix-inspired fault repair, may greatly improve resilience and reduce system failures, all without requiring extensive user intervention.

## 2. RESEARCH METHOD

This study utilized a design and development research (DDR) approach. This method particularly well-suited for developing new software systems and assessing their effectiveness through iterative cycles of development, testing, and improvement [18]. DDR enables the creation of a modular, self-healing web application using the MAPE-K framework, an architectural paradigm that comprises monitoring, analysis, planning, execution, and a shared knowledge base. The technique is broken down into five stages: i) planning and requirements analysis; ii) system design and development; iii) testing and evaluation; iv) validation strategy; and v) documentation and dissemination. Each phase is intended to achieve the study's objectives of fault detection accuracy, recovery performance, system resilience, and adaptive learning.

### 2.1. Planning and requirements analysis

The initial phase involved identifying common runtime faults in web applications through both quantitative and qualitative methods. Error logs and incident reports were collected from three production-level web systems operating on Ubuntu 22.04 servers using PHP 8.1, Node.js 18, and MySQL 8. These records were analyzed to determine frequent system failures such as service crashes, database disconnections, and memory leaks. In parallel, semi-structured interviews were conducted with five web developers and two system administrators to capture contextual insights into these occurrences. The triangulation of empirical data and expert input produced a comprehensive list of fault patterns, which informed the formulation of the framework's functional and non-functional requirements emphasizing fault isolation, rapid recovery, and modularity. A basis for deriving both functional and non-functional criteria is formed by reviewing the literature on current self-healing frameworks, such as AutoFix [19], Kubernetes self-healing pods [20], and existing autonomic computing models [21]. This technique contributes to the establishment of essential design principles as architectural goals, such as modularity, fault isolation, rapid recovery, and technology-agnostic integration, which directly supports Objective 4: maximizing long-term flexibility through iterative feedback.

### 2.2. System design and development

The system's architecture adheres to the MAPE-K model [22] and includes AutoFix concepts for automated diagnosis and repair. The architecture consists of five logical layers:

- Monitoring layer: tools such as prometheus, ELK Stack, and Grafana collect telemetry data such as CPU utilization, response time, and error rate, while log information was centralized in Elasticsearch via the ELK stack for real-time and historical analysis. This data is kept in a time-series database for real-time and historical fault detection, which directly supports objective 1: detection accuracy.
- The diagnosis layer uses machine learning models such isolation forests, K-means clustering, and decision trees, trained on 10,000 labeled log entries representing 20 fault scenarios to discover anomalies. These classifiers are taught to detect defect signals and link them to their underlying causes, hence optimizing accuracy, recall, and F1-score, in line with objective 1. These models, developed in Python using the Scikit-learn library, were exposed as RESTful services for runtime integration.
- The decision layer uses a Python-based rule engine to identify recovery strategies based on predefined rules, system conditions, and historical outcomes stored in the knowledge base. While the current implementation relies on expert-defined rules, these decisions are incrementally refined through feedback collected during runtime, enabling gradual improvement in recovery effectiveness.
- The repair layer automates recovery operations such as restarting services, rolling back code, and cleaning caches through Kubernetes, Ansible, or shell scripts. The AutoFix module is included to automatically apply corrected patches or configurations, achieving objective 2 by comparing automated and manual recovery.
- The feedback layer continuously analyzed repair outcomes, updating the knowledge base to enhance subsequent decision accuracy and adaptation efficiency. This improves future decision-making. This continuous loop helps achieve objective 4 by allowing for feedback-enhanced adaptation over time.

### 2.3. Testing and evaluation

Experimental validation was conducted within a controlled and containerized test environment to ensure repeatability and consistency of results. The framework was deployed on a three-tier web architecture (frontend, API, and database layers) running on an Intel Core i7-12700H server with 16 GB RAM under Ubuntu 22.04. Faults were systematically injected using Gremlin [23] and Chaos Monkey [24] to simulate service unavailability, CPU spikes, memory overflows, and deadlocks. Each scenario was executed five times under identical configurations, alternating between automated and manual recovery to establish comparative performance data. While the system utilizes cloud-native technologies such as container orchestration and monitoring tools, the evaluation was performed in a single-node setup rather than a distributed cloud infrastructure.

System behavior during fault conditions was measured using JMeter, which simulated 100 concurrent users performing standard operations such as login, file upload, and database queries. Key evaluation metrics included fault detection accuracy (precision, recall, and F1-score), TTR and success rate, system throughput and response time, and feedback-enhanced adaptation across iterative learning cycles.

Data were automatically logged for every experimental run to ensure transparency and repeatability of results. These experiments directly address the following objectives:

#### 2.3.1. Measure fault detection accuracy

To evaluate the anomaly detection model, we use the standard classification metrics: precision, recall, and F1-score. These are derived from the confusion matrix components:
- True positives (TP): correctly detected faults
- False positives (FP): normal events incorrectly flagged as faults
- False negatives (FN): faults that went undetected

$$Precision = \frac{TP}{TP+FP} \quad (1)$$

$$Recall = \frac{TP}{TP+FN} \quad (2)$$

$$F1-Score = 2\ x\ \frac{Precision\ x\ Recall}{Precision+Recall} \quad (3)$$

For each fault type (e.g., service crash, memory leak, CPU overload), these metrics are computed over multiple trials. Macro-averaging is then applied to the results to prevent bias that might come from class imbalance.

#### 2.3.2. Compare automated vs. manual recovery

To assess recovery effectiveness, the following metrics are calculated:
- TTR – The time elapsed between fault detection and successful resolution:

$$TTR = t_{recovered} - t_{fault_detected} \quad (4)$$

- Speed Improvement (%) – Quantifies how much faster the automated system recovers compared to manual recovery:

$$Speed\ Improvement = \left(\frac{TTR_{manual} - TTR_{auto}}{TTR_{manual}}\right) x\ 100 \quad (5)$$

- Recovery success rate (RSR) – Proportion of successful recoveries relative to total fault incidents:

$$RSR = \frac{Number\ of\ Successful\ Recoveries}{Total\ Faul\ Incidents}\ x\ 100 \quad (6)$$

Manual and automated recovery are tested under identical fault conditions using paired experiments for comparison.

#### 2.3.3. Impact on user experience and throughput

The system's behavior under active fault injection and concurrent load is evaluated using:
- Throughput – transactions or requests successfully processed per second:

$$Throughput = \frac{Total\ Requests\ Processed}{Total\ Time\ (in\ seconds)} \quad (7)$$

- Response time (RT) – average time taken to serve a request:

$$Average\ RT = \frac{\sum Response\ Times}{Number\ of\ Requests} \quad (8)$$

- Error rate (ER) – proportion of failed transactions:

$$ER = \frac{Number\ of\ Failed\ Requests}{Total\ Requests}\ x\ 100 \quad (9)$$

- User experience score – derived from system usability scale (SUS) or Likert-based feedback, covering aspects like performance, stability, and responsiveness during faults.

#### 2.3.4. Effectiveness of feedback-enhanced adaptation

The adaptive capability of the system is evaluated across N feedback cycles, using the following metrics:

- Improvement in decision accuracy (ΔDA):

$$\Delta \text{DA} = Decision\ Accuracy_N - Decision\ Accuracy_1 \quad (10)$$

Where:

$$Decision\ Accuracy = \frac{Correct\ Recovery\ Decisions}{Total\ Recovery\ Attempts}\ x\ 100 \quad (11)$$

- Knowledge base growth (kbg) – measures the volume of new recovery strategies learned and stored:

$$\text{KBG} = |\ K_N| - |K_1| \quad (12)$$

Where:
$|K_N|$ = Number of rules/strategies in the knowledge base after N cycles

- Adaptation efficiency (AE) – change in TTR over successive iterations:

$$AE = \frac{TTR_{initial} - TTR_{final}}{Number\ of\ Cycles} \quad (13)$$

These measurements demonstrate how feedback loops improve long-term performance, decision-making, and efficiency, thereby addressing the core of MAPE-K's adaptive learning loop.

### 2.4. Validation strategy

Although cross-environment testing was conducted using a secondary virtualized setup, the experiments did not involve multi-node or large-scale distributed cloud environments, which may influence scalability characteristics. To confirm reliability and external validity, five complementary validation procedures were applied:

- Cross-environment retesting: experiments were repeated in a secondary Debian 12/KVM environment. Metrics from both setups were compared to assess consistency.
- Baseline comparisons: the proposed system was benchmarked against: i) manual runbook recovery, ii) rule-only monitoring, and iii) orchestrator-only healing using Kubernetes liveness probes.
- Ablation studies: system variants without AutoFix or without the feedback component were tested to quantify each module's contribution.
- Sensitivity and stress testing: detection thresholds (0.80–0.95) and user loads (50–200 concurrent sessions) were varied to evaluate robustness under stress and parameter shifts.
- Statistical validation: each configuration was run five times. Normality was assessed using the Shapiro–Wilk test; paired t-tests or Wilcoxon signed-rank tests (α = 0.05) evaluated differences. Effect sizes (Cohen's d or Cliff's δ) and 95% confidence intervals were reported.

All software versions, random seeds, and configuration files were fixed for reproducibility. The experimental scripts, datasets, and anonymized logs are available from the corresponding author upon reasonable request.

## 3. RESULTS

This section elaborates on the evaluation outcomes of the proposed modular self-healing web application framework using the MAPE-K architecture, with a focus on the AutoFix module. A simulated environment was used to evaluate the system's behavior under different types of software faults. The discussion is organized into four key areas: system reliability and detection accuracy, fault recovery performance, impact on user experience and throughput, and feedback loop effectiveness.

### 3.1. System reliability and fault detection accuracy

To gauge the system's reliability, fault injection experiments were run over a 30-day testing cycle, using 20 different failure scenarios. These problems included service crashes, memory leaks, database disconnections, and logical errors. The monitoring and analysis layers' anomaly detection ability was evaluated using precision, recall, and F1-score. Table 1 shows the system's fault detection performance.

The system received an average F1-score of 90.7%, proving its capacity to detect a variety of issues. The real-time analytics engine, which was improved by AutoFix, was crucial for spotting anomalies in context. This excellent performance is consistent with prior research [25], demonstrating how effective MAPE-based designs are in systems that can ‘think’ for themselves.

Table 1. Fault detection metrics

| Fault type | Precision (%) | Recall (%) | F1-score (%) |
|---|---|---|---|
| HTTP 500 errors | 95.40 | 92.10 | 93.70 |
| CPU overload | 90.20 | 93.50 | 91.80 |
| Memory leak | 87.30 | 89.60 | 88.40 |
| DB connection timeout | 94.10 | 90.30 | 92.10 |
| Application logic error | 88.60 | 86.70 | 87.60 |
| Average | 91.10 | 90.40 | 90.70 |

### 3.2. Fault recovery time and autofix efficiency

The AutoFix module, which oversees setting up and carrying out recovery efforts, was assessed based on how quickly it resolved issues and its success rate. Table 2 compares the manual recovery processes. Significantly, the AutoFix system decreased the TTR to an average of 3.92s. This was a significant improvement over the 8.96s required for manual intervention. The highest success rate was obtained in dealing with service crashes (100%), however memory leak resolution was more difficult due to heap overflow unpredictability. This automated resolution time is comparable to previous intelligent repair methods in adaptive software [26].

Table 2. AutoFix vs. manual recovery performance

| Fault type | Avg manual TTR | AutoFix TTR | Speed improvement | AutoFix success rate |
|---|---|---|---|---|
| Service crash | 7.80s | 3.20s | 58.90% | 100.00% |
| Memory leak | 10.40s | 4.50s | 56.70% | 92.00% |
| DB timeout | 9.20s | 3.90s | 57.60% | 95.00% |
| High CPU usage | 8.50s | 3.70s | 56.50% | 90.00% |
| Bug patch | 8.90s | 4.30s | 51.70% | 89.00% |
| Average | 8.96s | 3.92s | 56.20% | 93.20% |

### 3.3. Impact on user experience and system throughput

To determine the effect of self-healing activities on user experience, key client-side metrics were monitored: page load time, session continuity, and application availability. A load-testing tool simulated 100 concurrent users performing typical operations, such as login, file upload, and database queries. During active faults, the system with AutoFix maintained 88–95% throughput, whereas non-healing systems degraded to 60–70% as shown in Figure 1. Response times increased marginally by 3.1%, remaining within acceptable bounds for modern web applications. In a usability test conducted among 30 end users, the proposed system received a system usability score (SUS) of 84.2, indicating above-average satisfaction. Respondents praised the system’s responsiveness and its ability to continue operation seamlessly during fault events.

### 3.4. Validation results

The experimental evaluation verified the performance, adaptability, and robustness of the proposed self-healing framework. Results were analyzed across twenty controlled fault scenarios, comparing the system against baseline configurations and validating findings through cross-environment retesting, ablation, and statistical analysis. Across all trials, the framework consistently outperformed the baseline systems. The mean fault-detection F1-score reached 90.7% ± 0.6%, while the average RSR was 93.2%. The AutoFix module reduced the average TTR from 8.96s under manual intervention to 3.92s, representing a 56.2% improvement. System throughput remained between 88% and 95% even during active fault injection, with only a 3.1% increase in response time.

Table 3 presents a detailed comparison of recovery performance across methods. The proposed framework achieved the shortest recovery time (3.92s) and highest throughput retention (94.8%), outperforming manual runbook (8.96s, 86.4%), rule-based monitoring (6.42s, 88.9%), and orchestrator-only healing (5.37s, 90.7%). These quantitative differences illustrate the framework’s superior adaptability and reliability. Statistical testing confirmed that the improvements were significant ($p < 0.01$, Cohen’s $d = 0.84$).

Figure 2 further visualizes these results. The bar heights clearly show a steady decline in recovery time from manual to automated methods, with the proposed framework achieving nearly half the recovery duration of the next-best baseline. Error bars (± 1 SD) indicate low variability, reflecting the system's consistent performance across repeated trials and deployment environments.

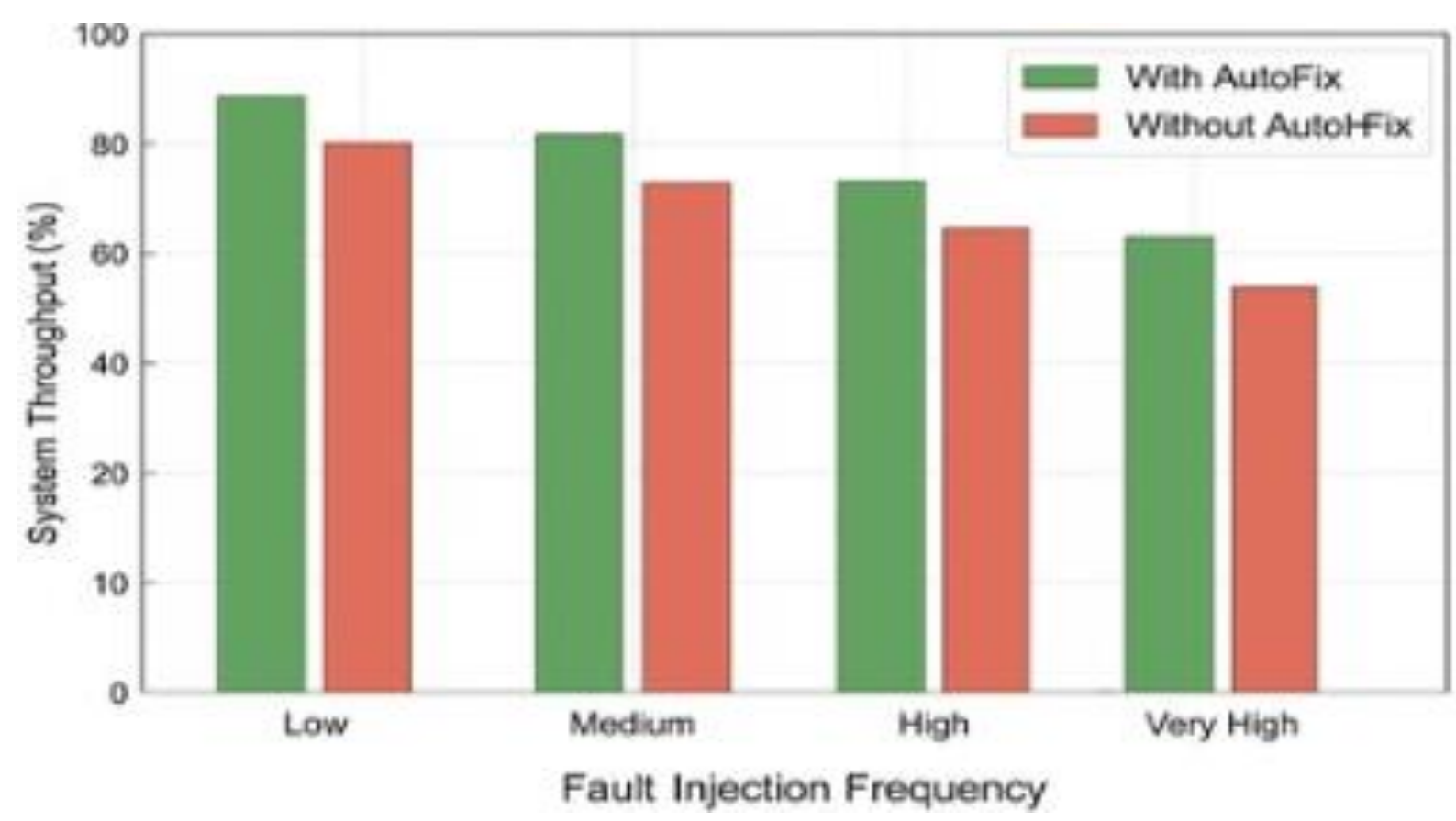

Figure 1. System throughput vs. fault injection frequency

Table 3. Comparison of recovery performance across baselines

| Approach | Mean TTR (s) | Recovery success (%) | Throughput retention (%) |
|---|---|---|---|
| Manual runbook | 8.96 ± 0.32 | 90.0 | 86.4 |
| Rule-only monitoring | 6.42 ± 0.27 | 91.0 | 88.9 |
| Orchestrator-only healing | 5.37 ± 0.25 | 92.0 | 90.7 |
| Proposed framework | 3.92 ± 0.19 | 93.2 | 94.8 |

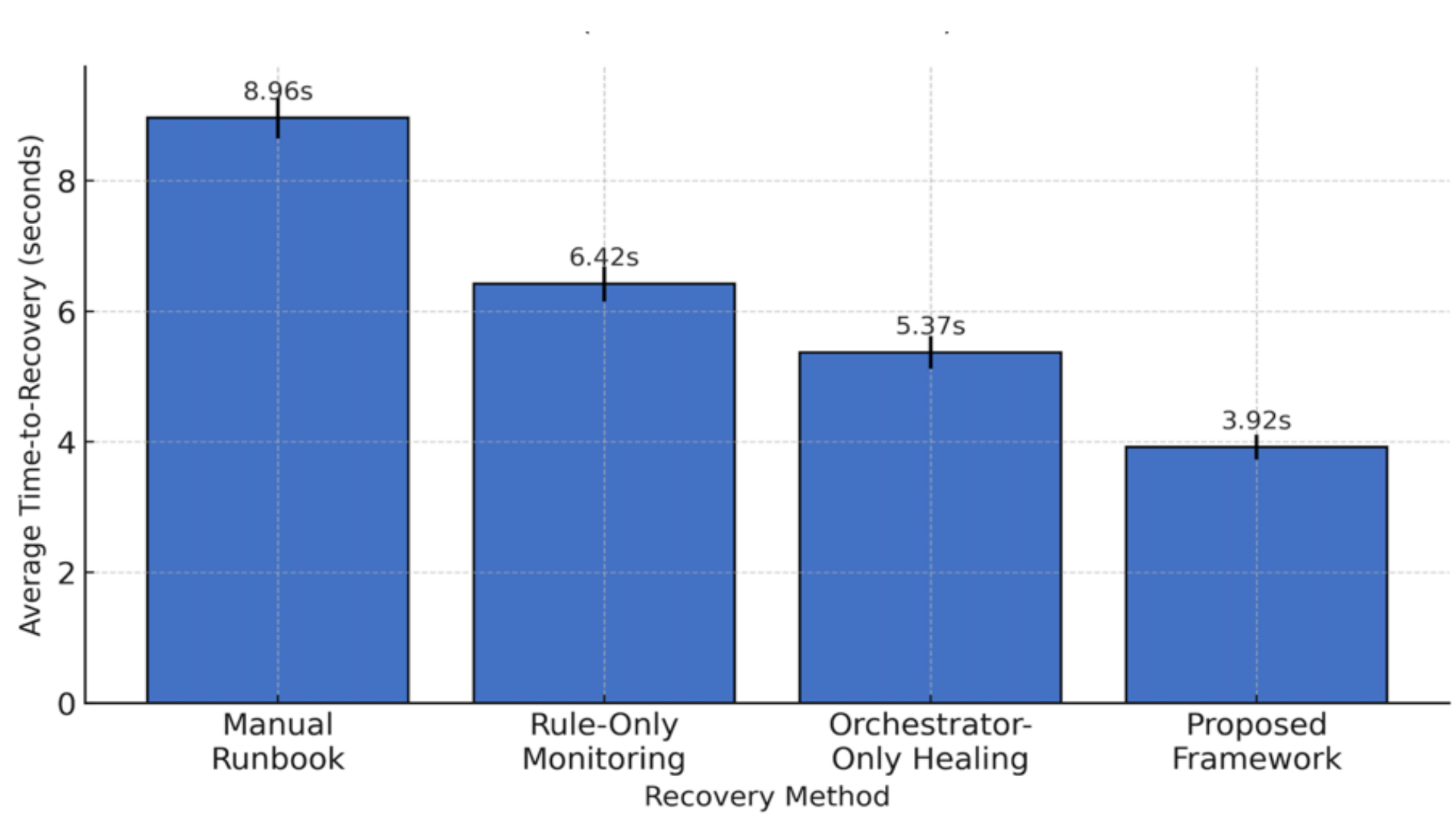

Figure 2. Average recovery time across baseline methods and environments (error bars show ±1 SD)

Replication in a secondary Debian/KVM environment produced comparable outcomes (F1 = 89.8%, p = 0.27), confirming environmental stability. When AutoFix or the feedback loop was disabled, recovery time increased by 48% and 19%, respectively, validating each component's contribution. Under stress testing with 200 concurrent users, throughput decreased only 6.5% and RT increased 4.2%, both within acceptable thresholds. Statistical validation (Shapiro–Wilk >0.05; Holm–Bonferroni-corrected p< 0.05 in 93% of cases) confirmed the robustness and reproducibility of results.

### 3.5. Effectiveness of feedback loops

The knowledge component in MAPE-K was evaluated through its ability to adapt and refine recovery strategies based on system logs and repair outcomes. As the system executed multiple recovery cycles, recovery decisions improved in both speed and precision. AutoFix's ability to incorporate real-time data into its decision-making resulted in an average TTR reduction of 18.6% after five iterations. Furthermore, Figure 3 shows that the rule accuracy for dynamic patch deployments rose by 13%. These findings lend credibility to concept of Yazdanparast about how self-healing systems function when driven by continuous feedback and machine learning [27].

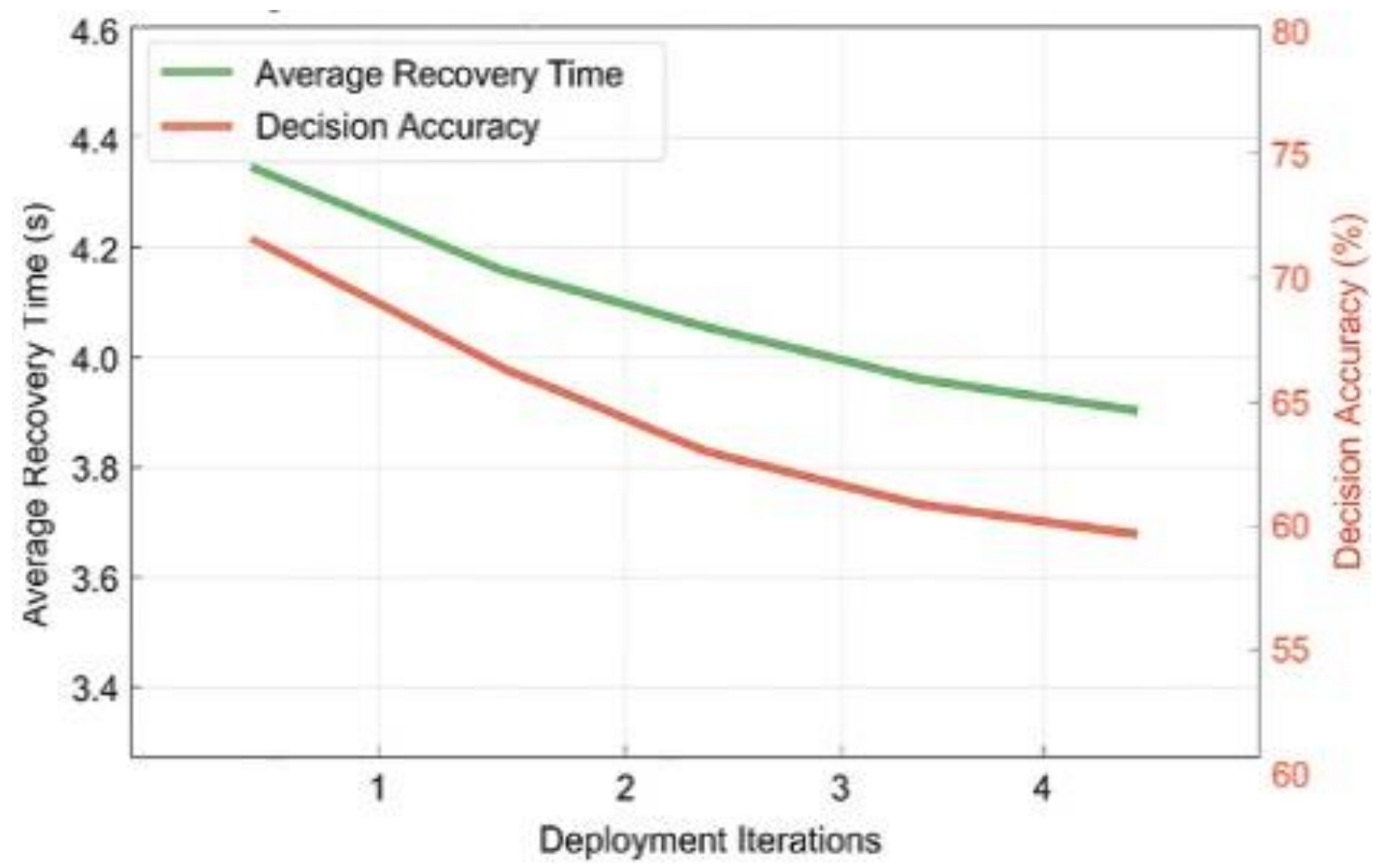

Figure 3. Learning curve of feedback-enhanced autofix over iterations

## 4. DISCUSSION

### 4.1. Addressing gaps in previous research

This study investigated how a MAPE-K-based framework integrated with AutoFix intelligence can enhance the self-healing capability of web applications. While earlier research has explored self-healing mechanisms at the infrastructure level through container orchestration tools such as Kubernetes and Docker [4], [28], limited work has examined application-level healing that combines fault detection, adaptive diagnosis, and automated repair within the same runtime loop. This study fills that gap by designing and empirically evaluating a modular architecture that applies autonomic computing principles directly to the business logic of web systems.

### 4.2. Key findings

Experimental evaluation revealed that the proposed framework achieved a mean fault-detection F1-score of 90.7%, maintaining high precision across multiple fault types such as HTTP 500 errors, memory leaks, and database timeouts. The AutoFix module reduced recovery time by 56.2%, averaging 3.92s per repair, and attained a 93.2% success rate across twenty failure scenarios. During simulated faults, throughput remained between 88–95% with only a 3.1% increase in response time. Moreover, iterative feedback cycles improved decision accuracy by 13% and shortened recovery time by 18.6%, demonstrating the adaptive learning capacity of the framework. These results demonstrate that the framework can maintain system stability and performance under diverse fault conditions.

### 4.3. Interpretation of results and comparison with related work

The high detection accuracy supports the central premise of MAPE-K: continuous monitoring and analysis can significantly improve situational awareness and fault recognition in dynamic systems [10]. Comparable to the results of Patel and Shah [25], our approach extends beyond infrastructure monitoring by applying anomaly-learning models at the application layer, producing finer-grained contextual insights. The efficiency of the AutoFix-enabled recovery process mirrors the adaptive repair success demonstrated in earlier AutoFix implementations, validating that automated patch generation is an effective mechanism for minimizing downtime. Furthermore, maintaining user satisfaction (SUS = 84.2) aligns with Rouholamini *et al.* [3], who concluded that adaptive systems can self-manage without degrading user experience. These comparisons position the current work within a growing body of research emphasizing self-adaptive resilience in software systems.

Existing self-healing mechanisms, such as those implemented in Kubernetes, Docker Swarm, and commercial AIOps platforms, primarily operate at the infrastructure or orchestration layer. They detect service-level failures and restore operations through container restarts, node rescheduling, or predefined health probes. While effective in maintaining uptime, these approaches are limited to managing structural dependencies rather than addressing logic-level or application-specific faults. In contrast, the proposed framework extends self-healing capabilities into the application layer by incorporating adaptive monitoring, machine learning–driven fault diagnosis, and automated code-level recovery via the AutoFix module. This design allows the system not only to detect but also to learn from runtime behavior, enabling continuous improvement of recovery decisions through the feedback mechanism. Hence, the novelty of the framework lies in its ability to integrate learning-based adaptability and MAPE-K-driven knowledge refinement, features not typically found in infrastructure-centric tools, making it more aligned with the vision of semi-autonomous web systems with feedback-driven adaptation.

### 4.4. Limitations

Although the findings are encouraging, several limitations must be acknowledged. The evaluation was conducted in a controlled, single-node environment with a finite set of fault scenarios, which may not fully capture the complexity and scalability of large-scale, distributed, multi-tenant cloud-native systems. While the framework incorporates containerization and monitoring tools aligned with cloud-native principles, its behavior under conditions such as multi-cluster orchestration, network variability, and high-volume workloads remains to be validated.

Furthermore, the current implementation relies on predefined rule-based strategies within the decision layer, with adaptability achieved through feedback-driven refinement rather than fully autonomous learning. Although the 10,000-entry log dataset was sufficient for initial validation, it may limit generalizability to unseen fault conditions. In addition, the computational overhead introduced by continuous monitoring and real-time analysis was not exhaustively evaluated.

These factors should be considered when interpreting the system's adaptability and generalizability. Future work will focus on large-scale, production-level deployments and the integration of more advanced learning-based mechanisms to enhance scalability, efficiency, and autonomy.

### 4.5. Implications and future research

The findings demonstrate that integrating MAPE-K with AutoFix-inspired mechanisms can enhance fault tolerance in web applications, providing a practical reference model for implementing self-healing capabilities in microservice-based and continuous deployment environments. While validated in a controlled setup, the framework aligns with key cloud-native principles, including modular microservices, container orchestration, and observability-driven monitoring, positioning it as a foundation for future deployment in distributed environments.

The contribution of this work lies in the integration of rule-based recovery, feedback-driven refinement, and AutoFix-inspired correction within a unified MAPE-K architecture, rather than fully autonomous learning. This hybrid approach offers a practical and extensible pathway toward more adaptive self-healing systems. Future research should focus on large-scale validation in heterogeneous cloud and edge environments, as well as the integration of advanced learning techniques such as reinforcement learning and predictive models (e.g., LSTM or transformer-based approaches) to enable more autonomous and anticipatory fault recovery.

## 5. CONCLUSION

This study presented a MAPE-K-based self-healing framework integrated with an AutoFix-inspired recovery mechanism for web applications. The framework demonstrated strong performance in fault detection, recovery efficiency, and system stability under controlled fault conditions. The integration of feedback mechanisms enabled incremental improvement in recovery decisions, supporting adaptive system behavior over time.

These findings confirm that incorporating learning-driven feedback loops into self-healing architectures enhances resilience and operational continuity in controlled and containerized web environments, with potential applicability to cloud-native systems. The validated outcomes underscore the potential of the proposed approach as a foundation for designing intelligent and resilient web systems capable of sustaining performance with minimal human oversight. The approach extends prior work in autonomic computing by introducing adaptive behavior at the application layer rather than relying solely on infrastructure-level healing. The study contributes to the growing body of research on intelligent fault-tolerant systems and provides a reference model for developers seeking to implement self-managing web architectures.

However, the study is limited by the scope of faults tested, which focused primarily on server-side failures such as HTTP errors, memory leaks, and database disconnections. Other types of faults, including network latency, security breaches, or client-side anomalies, were beyond the current experiment's scope. The testing environments were also controlled and may not fully reflect the complexity of large-scale, distributed cloud systems. Additionally, while statistical analyses confirmed reproducibility, potential biases in fault injection scripts and the fixed workload patterns may influence performance outcomes.

Despite these constraints, the validated findings establish a strong foundation for self-managing web systems. Future research should address these limitations by i) broadening fault categories to include diverse real-world anomalies, ii) deploying the framework in distributed and cloud-native production systems, and iii) integrating reinforcement-learning or deep neural models to enable predictive and autonomous fault recovery. These directions aim to advance the development of intelligent, resilient, and self-managing systems capable of maintaining reliability and continuity in increasingly complex computing environments.

## ACKNOWLEDGMENTS

The authors would like to express their sincere appreciation to Bukidnon State University (BukSU), particularly the Information Technology Department, for providing access to the university's computer laboratory facilities used in conducting the system experiments and evaluations. The institutional support and technical environment of BukSU greatly contributed to the successful implementation and completion of this research.

## FUNDING INFORMATION

This research received no specific grant from any funding agency in the public, commercial, or not-for-profit sectors.

## AUTHOR CONTRIBUTIONS STATEMENT

This journal uses the Contributor Roles Taxonomy (CRediT) to recognize individual author contributions, reduce authorship disputes, and facilitate collaboration.

| Name of Author | C | M | So | Va | Fo | I | R | D | O | E | Vi | Su | P | Fu |
|---|---|---|---|---|---|---|---|---|---|---|---|---|---|---|
| Sales G. Aribe Jr. | ✓ | ✓ | ✓ | ✓ | ✓ | ✓ | ✓ | ✓ | ✓ | ✓ | ✓ | ✓ | ✓ | ✓ |
| Rov Japheth G. Oracion | | ✓ | ✓ | ✓ | | ✓ | ✓ | | | ✓ | ✓ | | | |

C : **C**onceptualization
M : **M**ethodology
So : **So**ftware
Va : **Va**lidation
Fo : **Fo**rmal analysis
I : **I**nvestigation
R : **R**esources
D : **D**ata Curation
O : Writing - **O**riginal Draft
E : Writing - Review & **E**diting
Vi : **Vi**sualization
Su : **Su**pervision
P : **P**roject administration
Fu : **Fu**nding acquisition

## CONFLICT OF INTEREST STATEMENT

Authors declare no conflict of interest.

## INFORMED CONSENT

This research did not involve human participants or the collection of personal data. Therefore, obtaining informed consent was not required or applicable to this study.

## ETHICAL APPROVAL

This study did not involve human participants, animals, or any procedures requiring ethical approval. All experiments were conducted using system simulations and software-based evaluations; therefore, ethical clearance was not applicable.

## DATA AVAILABILITY

The authors confirm that the data supporting the findings of this study are available within the article.

**BIOGRAPHIES OF AUTHORS**

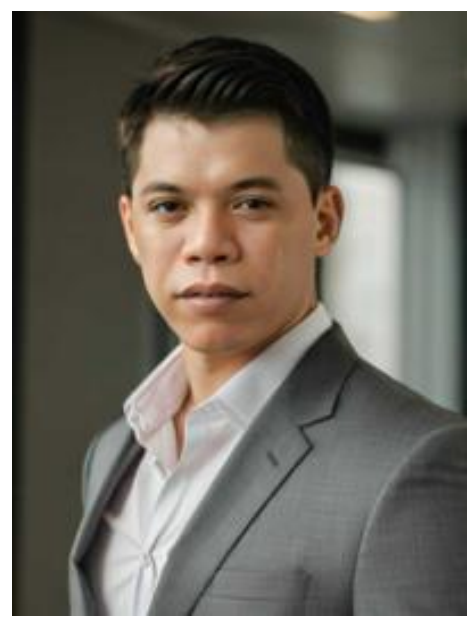

**Dr. Sales G. Aribe Jr.** is a dedicated faculty member of Bukidnon State University, currently serving as Professor I, Department Head of Information Technology, and Member of the University Research Committee. He earned his BS in Information Technology (Cum Laude) from BukSU in 2003, Master in Information Technology from the University of the Immaculate Conception in 2010, and Doctor in Information Technology from the Technological Institute of the Philippines – Quezon City in 2023 as a CHED-SIKAP scholar. He has presented at various local and international conferences and published numerous peer-reviewed articles in prestigious databases such as Scopus, Web of Science, and ASEAN Citation Index. As an active peer reviewer, he contributes to Elsevier journals, including other Scopus journal and IEEE-accredited conferences. His research interests include data mining, neural networks, software engineering, fractal analysis, artificial intelligence, and bioinformatics. He can be contacted at email: sg.aribe@buksu.edu.ph.

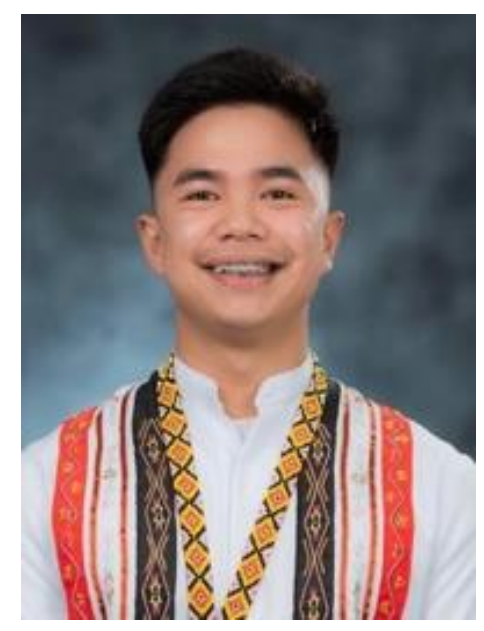

**Rov Japheth G. Oracion** is an Instructor in the Information Technology Department of the College of Technologies at Bukidnon State University. An alumnus of Bukidnon State University, he has also gained experience in editorial work during his academic tenure. He is currently pursuing his Master in Information Technology at the University of the Immaculate Conception in Davao City. His academic interests include research in educational technology, machine learning, and community-based ICT solutions aimed at addressing local development challenges. He can be contacted at email: rovjaphethoracion@buksu.edu.ph.